\documentclass[aps,prl,showpacs,twocolumn,superscriptaddress]{revtex4-1}
\usepackage{amsmath}
\usepackage{amsfonts}
\usepackage{braket}
\usepackage{amssymb}
\usepackage{graphicx}
\usepackage{natbib}
\usepackage{comment}
\usepackage[colorlinks=true,linkcolor=blue,urlcolor=black,citecolor=blue]{hyperref}
\usepackage{textcomp}
\usepackage{soul}
\usepackage[usenames,dvipsnames]{xcolor}

\newcommand{\ra}{\rangle}

\newcommand{\ti}{\tilde}

\begin{document}
\title{Expedited Holonomic Quantum Computation via Net Zero-Energy-Cost Control in Decoherence-Free Subspace}

\author{P. V. Pyshkin}
\affiliation{Beijing Computational Science Research Center, Beijing 100084, China}
\affiliation{Department of Theoretical Physics and History of Science, The Basque Country University (EHU/UPV), PO Box 644, 48080 Bilbao, Spain}
\affiliation{Ikerbasque, Basque Foundation for Science, 48011 Bilbao, Spain}

\author{Da-Wei Luo}
\affiliation{Beijing Computational Science Research Center, Beijing 100084, China}
\affiliation{Department of Theoretical Physics and History of Science, The Basque Country University (EHU/UPV), PO Box 644, 48080 Bilbao, Spain}
\affiliation{Ikerbasque, Basque Foundation for Science, 48011 Bilbao, Spain}

\author{Jun Jing}
\affiliation{Institute of Atomic and Molecular Physics and Jilin Provincial Key Laboratory of Applied Atomic and Molecular Spectroscopy, Jilin University, Changchun 130012, Jilin, China}
\affiliation{Department of Theoretical Physics and History of Science, The Basque Country University (EHU/UPV), PO Box 644, 48080 Bilbao, Spain}

\author{J. Q. You}
\affiliation{Beijing Computational Science Research Center, Beijing 100084, China}

\author{Lian-Ao Wu}
\thanks{To whom all correspondence should be addressed. lianao.wu@ehu.es}
\affiliation{Department of Theoretical Physics and History of Science, The Basque Country University (EHU/UPV), PO Box 644, 48080 Bilbao, Spain}
\affiliation{Ikerbasque, Basque Foundation for Science, 48011 Bilbao, Spain}

\date{\today}

\begin{abstract}
Holonomic quantum computation (HQC) may not show its full potential in quantum speedup due to the prerequisite of a long coherent runtime imposed by the adiabatic condition. Here we show that the conventional HQC can be dramatically accelerated by using external control fields, of which the effectiveness is exclusively determined by the integral of the control fields in the time domain. Remarkably this control scheme can be realized with net zero energy cost and it is fault-tolerant against fluctuation and noise, significantly relaxing the experimental constraints. We demonstrate how to realize the scheme via decoherence-free subspaces. In this way we unify quantum robustness merits of this fault-tolerant control scheme, the conventional HQC and  decoherence-free subspace, and propose an {\em expedited} holonomic quantum computation protocol.
\end{abstract}

\pacs{03.65.-w, 42.50.Dv, 37.10.De}


\maketitle

{\it Introduction.}{\bf---} As building blocks for quantum computers, the implementation of quantum gates has received considerable research efforts over the recent years~\cite{Nielsen2000}. It has been reported experimentally that numbers of pulse-controlled microscopic systems, such as solid-state spins~\cite{Arroyo-Camejo2014} and trapped ions~\cite{nat_cnot}, can be hosts for implementation of quantum gates. While enormous theoretical strategies for conventional quantum gate implementation have been proposed, there is a revived interest in using geometric phases to perform circuit-based quantum computation, termed as holonomic quantum computation (HQC)~\cite{Zanardi1999} , which is enabled by the adiabatic quantum theorem. The theorem asserts that at any instant a quantum system remains nearby in its instantaneous eigenstate of a slow-varying Hamiltonian, specifically for a cyclic adiabatic process, a geometric phase (the Berry's phase), is acquired over the course of the cycle~\cite{Berry1984}. The geometric phase is exclusively determined by the trajectory of the system in its parameter space and robust against local fluctuation~\cite{DiVincenzo2000,Carollo2005}. Consequently, a geometric strategy for implementation of quantum gates permits fault-tolerant and robust quantum information processing. Besides inherent resilience in non-Abelian geometric phases~\cite{Wilczek1984}, HQC has an appealing advantage~\cite{Niskanen2003,Oreshkov2009,Sjoeqvist2012} in utilizing the state-of-art experimental setups due to its close relationship to the circuit model~\cite{Deutsch1989,Mizel2007,Siu2005}. A recent experiment has implemented a universal set of geometric quantum logic gates with diamond nitrogen-vacancy centers~\cite{lmd}, and evidently it will greatly promote research endeavour along this line.

The heart of HQC is the experimental implementation of the geometric phase acquired in a cyclic adiabatic passage. Despite its advantages, the geometric protocol itself is challenged with a dilemma. On one hand, any HQC algorithm requires a long characteristic runtime in order to satisfy the adiabatic condition~\cite{Born1928}. On the other hand, decoherence or leakage accumulated in this long runtime gives rise to errors in the HQC processing and may eventually destroy the quantumness of the system. To get rid of the dilemma, researchers have proposed several different protocols. Over a decade ago, Wu, Zanardi and Lidar~\cite{Wu2005} initiated a scheme by embedding HQC into a decoherence-free subspace (DFS). This combined HQC-DFS scheme utilizes the virtues of both the fault-tolerance of HQC and the robustness of DFS against collective dephasing noise based on the symmetry structure of the interaction between the system and its environment. However, the residual individual noise remains and ruins the quantum adiabatic passages during the long runtime. Later on the HQC-DFS scheme was extended by considering the collective dephasing of two neighboring physical qubits~\cite{Xun-Li}. Whereas it is more feasible experimentally, this scheme has a more stringent requirement for the runtime. Recently a non-adiabatic HQC-DFS scheme was suggested where the characteristic timescale is reduced by increasing the characteristic energy, at the cost of a harsh restriction for the runtime equal to the period of the system~\cite{Tong2012}. However, the fault tolerance from adiabaticity therefore becomes obscure.

In this Letter, we propose a novel and composite strategy to tackle the long runtime issue in the HQC protocols via accelerating the adiabatic passage in~DFS. We explain the mechanism and show specifically that the characteristic timescale of the adiabatic process can be vastly reduced by means of external field control~\cite{Jing2014}. Interestingly, it is found that the particular design or shape of a control function, such as regular, random, chaotic and even noisy pulse sequences, is not as decisive as it seems to be, but only the integral of the control function in the time domain plays the crucial role in speeding up the adiabatic passage, which greatly relaxes constraints on experimental implementation of these control functions. Remarkably, we further discover that our Hamiltonians in the adiabatic representation are periodical functionals of the integral of the control functions, resulting in a  {\em net zero-energy-cost control scheme} -- a new mechanism that accelerates adiabatic passages with the same effectiveness. These lead to a new type of fault-tolerance against control fluctuations.

{\it Decoherence-free subspace for qubit gates.}{\bf---} Decoherence-free subspace is based on the symmetry structure of the system-environment interaction~\cite{dfs_ref1,dfs_ref2_1,dfs_ref2_2,dfs_ref2_3,dfs_ref2_4}. Here we briefly recall the method to realize a universal set of quantum gates acting on the DFS as firstly proposed in Ref.~\cite{Wu2005}. To implement a one-qubit quantum gate in DFS, we consider a four physical qubit system with the Hamiltonian $H = \sum_{l<m}(J_{lm}^xR_{lm}^x + J_{lm}^yR_{lm}^y)$, where $R_{lm}^x = \frac{1}{2}(\sigma_l^x \sigma_m^x + \sigma_l^y \sigma_m^y)$, $R_{lm}^y = \frac{1}{2}(\sigma_l^x \sigma_m^y - \sigma_l^y \sigma_m^x)$ are the XY interactions and Dzialoshinski-Moriya terms, $\sigma_i^{x(y)}$ is Pauli X(Y) matrix acting on the $i$-th physical qubit and $m,l=1,2,3,4$. This Hamiltonian commutes with the operator $Z = \sum_{i=1}^{4}\sigma_i^z$, where $\sigma_i^z$ is a $Z$ Pauli matrix acting on $i$-th physical qubit. By setting $J_{12}^x=J_{12}\cos\varphi(t)$, where $\varphi(t)$ is specifically designed for HQC, $J_{12}^y=J_{12}\sin\varphi(t)$, $J_{13}^x=J_{13}$ and all other $J_{lm}^{x(y)}\equiv 0$, the Hamiltonian becomes
\begin{align}\label{eqn_h0_wu}
H(t)& = J_{13}R_{13}^x + J_{12}\left[\vphantom{1^1}\cos\varphi(t)R_{12}^x - \sin\varphi(t)R_{12}^y\right].
\end{align}
The bases for DFS have been identified as eigenvectors of $Z$~\cite{Wu2005}, as spanned by $\{ \ket{0}, \ket{1}, \ket{2}, \ket{3} \}$, where $\ket{0} = \ket{0001}$ and $\ket{1} = \ket{0010}$ constitute the two orthonormal states for a logical qubit and $\ket{2}=\ket{1000}$ and $\ket{3}=\ket{0100}$ serve as ancilla. This DFS scheme is robust against collective dephasing described by $Z\otimes B$, where $B$ is an arbitrary Hermitian bath operator. It is straightforwardly proven that in the DFS, the Hamiltonian~\eqref{eqn_h0_wu} can be rewritten as
\begin{align}\label{dfs_ham}
H_1(t)=&\sin\theta(t)(\ket{1}\bra{2}+\ket{2}\bra{1}) \nonumber \\ &+\cos\theta(t)(e^{-i\varphi(t)}\ket{3}\bra{2} + e^{i\varphi(t)}\ket{2}\bra{3}),
\end{align}
where $\theta(t) = \tan^{-1}(J_{13}/J_{12})$.

{\it Holonomic quantum computation in DFS.}{\bf---} Consider a quantum system whose dynamics is governed by a time-dependent Hamiltonian $H(t)$ with instantaneous eigenvectors $|E_n(t) \rangle$ and eigenvalues $E_n(t)$. The wave function $\ket{\psi(t)}$ satisfies the Schr\"{o}dinger equation and can be formally written as $\ket{\psi(t)} = \sum_n\psi_n(t)e^{i\phi_n(t)}\ket{E_n(t)}$, where $\phi_n(t)\equiv-\int_{0}^{t}E_n(s)ds$ is the dynamical phase. If the Hamiltonian varies adiabatically and there is a non-vanishing gap between the interested eigenvalues, the system will remain in the corresponding instantaneous eigenstate. Consequently, a Berry's phase is given when the system passes along a closed loop in the Hamiltonian parameter space, which is path-independent. Without loss of generality, one can consider a case where the system is initially at the~$n$-th ground state $|E_n \rangle$. It follows that in the adiabatic regime $\psi_n=e^{i\gamma_n(t)}$, where $\gamma_n(t)$ is the Berry's phase given by $\gamma_n(t)=i\int_{0}^{t}\braket{E_n(s)|\dot{E}_n(s)}ds$. Here we emphasize that for dark states with eigenenergy $E_n(t)=0$, its dynamical phase vanishes and the remaining overall phase is a geometric phase.

Equipped with Eq.~(\ref{dfs_ham}), we are ready to construct {our expedited-HQC-DFS scheme}. To build up a one-qubit gate in DFS, we consider a cyclic Hamiltonian with period of $T$. We first consider a single qubit phase gate. The Hamiltonian $H_1(t)$ is formally given by Eq.~\eqref{dfs_ham} regarding $\theta(t) = a\sin \frac{2\pi t}{T}$, $\varphi(t)=\frac{2\pi t}{T}$, where $a$ is a dimensionless undetermined coefficient. The two dark states in the DFS for Hamiltonian $H_1(t)$ read as $\ket{D_0(t)} = \ket{0}$ and $\ket{D_1(t)} = \cos\theta(t)\ket{1} - e^{-i\varphi(t)}\sin\theta(t)\ket{3}$, respectively.

In the adiabatic regime, under the unitary evolution $U(T)=\mathcal{T}\exp[-i\int_{0}^{T}dsH(s)]$ where $\mathcal T$ is time-ordering operator, the dark states $\ket{D_0}$ and $\ket{D_1}$ become
\begin{equation}\label{D_1_transform}
e^{i\gamma_0(T)}\ket{D_0(T)}, \quad
e^{i\gamma_1(T)}\ket{D_1(T)},
\end{equation}
respectively, where $\gamma_{j}(T)$ is the Berry's phase for $|D_j\rangle$, $j=0,1$. In this manner we achieve a one-qubit phase gate by $e^{i\gamma_0(T)}|D_0(T)\rangle\langle D_0(0)|+e^{i\gamma_1(T)}|D_1(T)\rangle\langle D_1(0)|$. Note that $|D_j(T)\ra=|D_j(0)\ra$. The gate can be expressed by a diagonal matrix as ${\rm diag}([e^{i\gamma_0(T)}, e^{i\gamma_1(T)}])$. The two Berry's phases for dark states are $\gamma_0(T) = 0$ and
\begin{equation}\label{berry_phase_defin}
\gamma_1(T) =\int_{0}^{T}\sin^2\theta(s)\frac{\partial \varphi(s)}{\partial s}ds=\pi[ 1 - J_0(2a) ],
\end{equation}
where $J_0(x)$ is a zero order Bessel function of the first kind, respectively.

This technique is also applicable in realization of a single $\sigma_x$ qubit gate. To build this gate, we implement the Hamiltonian in the same DFS yet spanned by $\{\ket{+}, \ket{-}, \ket{2}, \ket{3}\}$, where $\ket{\pm} \equiv  (\ket{0}\pm\ket{1})/\sqrt{2}$. It is written as
\begin{align}\label{H-xgate}
H_2(t) &= \sin\theta(t)\left( \ket{-}\bra{2} + \ket{2}\bra{-}  \right) \nonumber\\
&+\cos\theta(t)\left( e^{-i\varphi(t)}\ket{3}\bra{2} + e^{i\varphi(t)}\ket{2}\bra{3}  \right).
\end{align}
In this case, the new dark states are $\ket{D_0(t)} = \ket{+}$ and $\ket{D_1(t)} = \cos\theta(t)\ket{-}-\sin\theta(t)e^{-i\varphi(t)}\ket{3}$, respectively. The transformations of dark states under time evolution are still described by Eq.~(\ref{D_1_transform}), and the qubit gate reads,
\begin{equation}\label{x_gate_1}
e^{i\gamma_1/2}\begin{pmatrix}
\cos\gamma_1/2 & -i\sin\gamma_1/2 \\
-i\sin\gamma_1/2 & \cos\gamma_1/2
\end{pmatrix},
\end{equation}
which becomes the $\sigma_x$-gate when $\gamma_1(T)=\pi$. 

Now we turn to the two-qubit controlled-phase (C-Phase) gate in DFS. Since each logical qubit consists of four physical qubits, eight physical qubits are involved in implementing a two logical-qubit gate. Let us suppose that one can implement the Hamiltonian
\begin{multline}\label{dfs_ham_c-phase}
H_3(t) = \sin\theta(t)\left(\ket{1, 1}\bra{2, 1}+\ket{2, 1}\bra{1, 1}\right) + \\
\cos\theta(t)\left(e^{-i\varphi(t)}\ket{3, 1}\bra{2, 1} + e^{i\varphi(t)}\ket{2, 1}\bra{3, 1}\right).
\end{multline}
The four dark states of the Hamiltonian employed in implementing C-Phase gate are given by $\ket{D_0(t)} = \ket{0, 0}$,  $\ket{D_1(t)} = \ket{0, 1}$, $\ket{D_2(t)} = \ket{1, 0}$, $\ket{D_3(t)} = \cos\theta(t)\ket{1, 1} - e^{-i\varphi(t)}\sin\theta(t)\ket{3, 1}$, respectively.

Over a period $T$, the Hamiltonian~\eqref{dfs_ham_c-phase} drives these states into $\ket{D_0(0)}\rightarrow\ket{D_0(T)}$,
$\ket{D_1(0)}\rightarrow\ket{D_1(T)}$,
$\ket{D_2(0)}\rightarrow\ket{D_2(T)}$ and
$\ket{D_3(0)}\rightarrow e^{i\gamma_3(T)}\ket{D_3(T)}$, so that the two-qubit gate is ${\rm diag}([1, 1, 1, e^{i\gamma_3(T)}])$, where $\gamma_3(T)=\gamma_1(T)$ in Eq.~(\ref{berry_phase_defin}). Tuning the free parameter $a$, one can get an arbitrary phase gate at will, for example, $\gamma_3(T)=\pi$ requires $J_0(2a)=0$ at the first root $a=1.2024$.


{\it Control scheme.{\bf---}} We now come to the case where the Hamiltonian $H(t)$ is {\em not} in the adiabatic regime. Our scheme is to implement a control $c(t)$ upon the strength of the Hamiltonian such that~\cite{Jing2014,Wang2014}
\begin{equation}\label{H_modif}
H(t)\rightarrow\left[\vphantom{\frac{1}{1}} 1+c(t) \right]H(t).
\end{equation}
We first show that as long as the control is sufficiently fast and strong, the system evolution will behave in the same way as that in the adiabatic regime, specifically the wave function $\ket{\psi(t)}$ becomes proportional to an instantaneous eigenstate of $H(t)$. {It is interesting to note that this control scheme hardly depends on the details of $c(t)$ but its integral in the time domain, and is a new type of fault-tolerance against control fluctuations.} Consequently, the evolution of the corresponding dark states are shown to be a qualified workstation for HQC and this {\em induced} adiabaticity will be utilized to realized the {\em expedited} HQC in virtue of a fast modulation over Hamiltonian.

We emphasize that the results given in Eq.~(\ref{berry_phase_defin}) are invariant under the transformation~(\ref{H_modif}), which is one of key points of our proposal.

\begin{figure}[htbp]
	\begin{center}
		\includegraphics[width=6 cm]{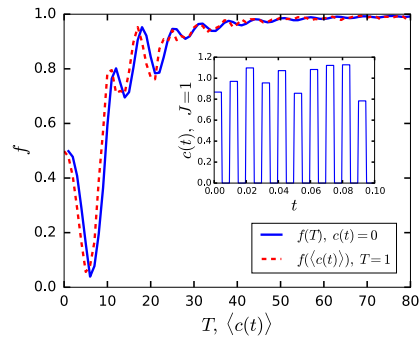}
	\end{center}
	\caption{(Color online) Quality factor $f$ as a function of the time $T$ as shown by the blue curve, where $T>60$ roughly corresponds to the conventional adiabatic condition for our model,  and as a function of the average control strength~$\langle c(t) \rangle$ for $T=1$ in the nonadibabtic domain.  The control $c(t)$ is modelled as a train of pseudo-periodical
	square pulses with a fixed period $2\Delta t$ and duty cycle $50\%$ (see example of~$c(t)$ in inset). 
	The amplitude of the control pulses is given by $J(1-p(1/2-r))$, where $r\in [0,1)$ is a uniform random number, $J>0$ is a parameter, and $p$ describes the randomness of the control (we use $p=0.5$ and $\Delta t = 0.005$). The Berry phase is numerically calculated by $e^{i\gamma_1}=\braket{D_1(0)|U(T)|D_1(0)}$. Here we used~$a=0.7605$ and $\gamma_1 = \pi/2$}. \label{fig_1}
\end{figure}

To determine the effectiveness of our control scheme, we now introduce a quality factor
$$f = \left(\vphantom{1^1}1 - \frac{|\delta\gamma_1|}{\pi}\right)\times\left|\vphantom{1^1}\braket{D_1(0)|U(T)|D_1(0)}\right|,$$
where $\delta \gamma_1$ is the difference between the ideal phase~\eqref{berry_phase_defin} and the phase acquired during a finite runtime $T$. Accordingly, we have $0\leq f\leq 1$ where $f=1$ if and only if the process simultaneously has ideal adiabaticity and retain the Berry phase predicted by~(\ref{berry_phase_defin}). Figure~\ref{fig_1} shows~$f$ as a function of evolution time~$T$ (blue curve) in the absence of control ($c(t)=0$), and as a function of {\em average} noise kick's strength~$\langle c(t)\rangle$ for~$T$ (red dashed curve) that is not in the adiabatic domain. 

{\it Mechanism of the adiabatic speedup.}{\bf---}To understand the mechanism of our expedited HQC scheme  we expand the wave function in terms of eigenstates~$\ket{E_n(t)}$ of the Hamiltonians. The matrix elements of the Hamiltonians in the adiabatic representation reads, $H_{mn}=\braket{E_n|\dot E_m}\exp{(i\int_{0}^{t}(1+c(t'))E_{mn}(t')dt')}$~\cite{Jing2014}. For example, the Hamiltonian~(\ref{dfs_ham}) is
\begin{widetext}
	\begin{equation}\label{H_rot_frame}
	\ti{H}_1(t) = \begin{pmatrix}
	0 & 0 & 0 & 0 \\
	0 & -\dot\varphi\sin^2\theta & (\dot{\theta}+\frac{i}{2}\dot{\varphi}\sin2\theta)e^{-iC(t)} & (\dot{\theta}+\frac{i}{2}\dot{\varphi}\sin2\theta)e^{iC(t)} \\
	0 & (\dot{\theta}-\frac{i}{2}\dot{\varphi}\sin2\theta)e^{iC(t)} & -\dot\varphi\cos^2\theta & -\dot\varphi\cos^2\theta e^{2iC(t)} \\
	0 & (\dot{\theta}-\frac{i}{2}\dot{\varphi}\sin2\theta)e^{-iC(t)} & -\dot\varphi\cos^2\theta e^{-2iC(t)} & -\dot\varphi\cos^2\theta
	\end{pmatrix},
	\end{equation}
\end{widetext}
where $C(t)=\int_{0}^{t}ds[1 + c(s)]$. It shows clearly that the Hamiltonian is a functional of the integral $C(t)$ (or the average of $c(t)$ in the time domain) i.e., $\ti{H}[C]$, meaning that controlled dynamics does not depend on the details of $c(t)$ but {\em exclusively depends} on the integral $C(t)$.  Such {\em exclusive dependence} also holds for our Hamiltonians~(\ref{dfs_ham}),(\ref{H-xgate}) and (\ref{dfs_ham_c-phase}), and is a unique feature of our chosen Hamiltonians whose energy differences $E_{nm}=E_{m} - E_{n}$ are time-independent constants. These Hamiltonians, as shown in its adiabatic representation, are incidentally equivalent to the Leakage Elimination Operators~\cite{Jing2015}. Hence, the control is fault tolerant in the sense that the fluctuation or noise of $c(t)$ hardly contributes to $C(t)$. More specially, by considering the propagator from $t=0$ to $t=\delta t$, where $\delta t \ll 1$ and $\braket{c(t)}\gg 1/\delta t \gg 1$, we can write the propagator as,
$U(\delta t)=\mathcal{T}\exp{\left(-i\int_{0}^{\delta t}\ti{H}_1(t)dt\right)}\approx 1 - i\int_{0}^{\delta t}\ti{H}_1(t)dt$. The existence of the fast oscillating factor $e^{iC(t)}$ renders all the off-diagonal elements of the propagator vanish and then leaves a Berry's phase to the amplitudes of $|D_1\ra$ and two bright eigenstates. Noticeably this factor pushes the evolution of system into the adiabatic regime by {\em decoupling} all the four eigenstates. It clearly illustrates the advantage of our control scheme: one needs not to care about the exact control function because only the integral $C(t)$ contributes to adiabaticity. 

\begin{figure}[htbp]
	\begin{center}
		\includegraphics[width=6 cm]{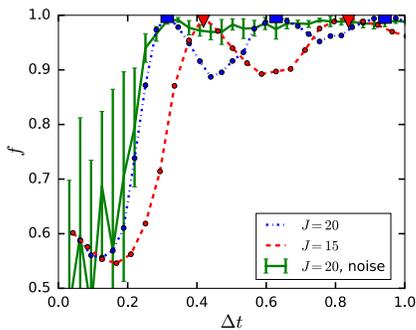}
	\end{center}
	\caption{(Color online) 
	Quality factor $f$ as a function of the kick length~$\Delta t$ for the zero-energy-cost control when $T=10$ in the nonadiabatic domain. We use control function $c(t) = J(1-p(1/2 -r))(-1)^{\lfloor t/\Delta t \rfloor}$, where $r$ is a random number. The blue dash-dotted and red dashed curves are for the noiseless control with $p=0$. The green solid curve represents noise control where each point is calculated with~$10$ random noise realizations where $p=0.5$. Triangles and squares signify points satisfying Eq.~(\ref{dtoscillations}). Here we used~$a=0.7605$ and $\gamma_1 = \pi/2$.}  \label{fig_2}
\end{figure}

{\it Expedited HQC with net zero energy cost.}{\bf---}  On closely looking into its pattern, we find that the Hamiltonian~(\ref{dfs_ham}) is eventually a functional of  the exponent $e^{iC(t)}$, i. e., $\ti{H}[e^{iC(t)}]$. Because of the periodicity of $e^{iC(t)}$, our control scheme allows for an interesting case when~$\langle c'(t) \rangle=0$, where $c'(t)$ has alternating positive and negative values such that the net energy cost is zero. 
We first illustrate that the above-discussed {\em positive} control $c(t)$ (with $\langle c(t) \rangle\gg0$) can be exactly equivalent to {\em zero-energy-cost} control $c'(t)$, when $c(t) = \pi \sum_i\delta(t - \tau_i)$ 
with the integral $C(t)$, and $c'(t) = \pi \sum_i(-1)^i\delta(t - \tau_i)$ with $C'(t)$. It is easy to show that $\ti H[e^{iC(t)}] =  \ti H[e^{iC'(t)}]$ due to the periodicity of $e^{iC(t)}$, and $\langle c(t) \rangle=2\pi/\Delta\tau_i$ ($\Delta\tau_i=\tau_{i+1}-\tau_i$) but $\langle c'(t) \rangle=0$ 
for each two consecutive pulses. 
The random intervals $\Delta\tau_i$ are much shorter than $T$ in reality,  and ideally the net energy cost of the $c'(t)$ control sequence can be considered as zero when $\Delta\tau_i$ approaches zero.  We can also analyze the equivalence for the rectangular pulses sequences.  Based on the first order of Magnus expansion of~$U(\delta t)$ we can justify~\cite{notice_1} that if the single pulse strength $J\gg 1$, the off-diagonal terms in evolution~$U(\delta t)$ become zero when
\begin{equation}\label{dtoscillations}
J\Delta t = 2\pi n, \; n=1,2,3\dots .
\end{equation}
In Fig.~\ref{fig_2} we show the numerical simulation of the quality factor $f$  for fixed $T=10$ with net zero-energy-cost control as a function of control pulse length~$\Delta t$.
We mark with triangles and squares when Eq.~(\ref{dtoscillations}) is satisfied. The green solid curve in Fig.~\ref{fig_2} shows the zero-energy-cost {\em noise control} which is more robust against the control ``kick'' length~$\Delta t$, while noise positive control has prominent oscillatory dependence on~$\Delta t$ which requires a more accurate choice of~$\Delta t$~(and/or~$J$) according to Eq.~(\ref{dtoscillations}).
	
{\it Conclusion.}{\bf---}To cope with the long runtime issue in implementing adiabatic passages, we have introduced an expedited-HQC-DFS control scheme to accelerate the conventional HQC. In the adiabatic representation, we show explicitly that the integral of the external control in the time domain typically the control pulse sequences, rather than the details of the control functions, exclusively determines the efficiency of speeding-up the runtime, such that the scheme is robust against the stochastic errors in control.  More importantly, we further find that the Hamiltonian in the adiabatic representation is a periodical functional of the integral of the control. The periodicity motivates us to design a {\em net zero energy cost}  strategy for speedup which is also robust against control imperfections. These novel results are confirmed by numerical results. This observation greatly reduces the experimental constraints in generating precisely-shaped pulses and allows us to use even random pulse sequences. By combining the features of this scheme with a scalable {DFS}, our expedited HQC protocol brings together the four-fold advantages of all-geometrical HQC, decoherence-free subspace, zero-energy-cost control, and our fault tolerant scheme, a typical {\em scalable, fast and fault-tolerant} architecture. We therefore expect that this perfect theoretical protocol becomes an experimental practice.

{\em Acknowledgments.}---We acknowledge grant support from the Basque Government (grant IT472-10), the Spanish MICINN (No. FIS2012-36673-C03-03), the NBRPC No.~2014CB921401, the NSAF No. U1330201, the NSFC Nos. 11575071 and 91421102, and Science and Technology Development Program of Jilin Province of China (20150519021JH).



\begin{thebibliography}{99}
\bibitem{notice_1}

Expression~(\ref{dtoscillations}) could be easily derived by the following consideration. We again consider short time evolution~$U(\delta t)$, and now we set $\delta t = \Delta t\ll T$, where $\Delta t$ is control pulse length. Adiabaticity means that off-diagonal elements of the matrix $\int_t^{t+\Delta t}\ti{H}_1(t')dt'$ are zero. Each of these off-diagonal elements could be written as $\int_t^{t+\Delta t}\exp(\pm iJt')F(t')dt'$, here we assume that $J\gg1$ and $F(t)$ is a smooth function: $F(t)\approx F(t+\Delta t)$, sign $\pm$ corresponds to the positive and negative control pulses. We can then conclude that off-diagonal elements of $U(\Delta t)$ becomes zero when $J\Delta t = 2\pi n$  ($n=1,2,3,\dots$), i.e. we have $\Delta t$ equals to integer numbers of periods $2\pi/J$.
\end{thebibliography}
\end{document}